\documentclass[preprint,12pt]{elsarticle}
\usepackage{graphicx}%
\usepackage{amsmath,amssymb,amsfonts}%
\usepackage{amsthm}%
\usepackage{mathrsfs}%
\usepackage[title]{appendix}%
\usepackage[dvipsnames]{xcolor}
\usepackage{textcomp}%
\usepackage{manyfoot}%
\usepackage{booktabs}%
\usepackage{algorithm}%
\usepackage{algorithmicx}%
\usepackage{algpseudocode}%
\usepackage{listings}%
\usepackage{enumitem}

\usepackage{url}
\usepackage{adjustbox}

\usepackage{caption}
\usepackage{stfloats}
\usepackage{amsmath}

\usepackage{hyperref}

\newcommand{\mc}[1]{\textcolor{black}{#1}}%{magenta}{#1}}

\journal{.}

\begin{document}

\begin{frontmatter}

%\title{Z-Dip: a validated generalization of the Dip Test for data modality assessment}
\title{Z-Dip: a standardized measure for data modality assessment}

\author[inst1]{Edoardo Di Martino\corref{cor1}}
\ead{edoardo.dimartino@uniroma1.it}
\author[inst2]{Matteo Cinelli}
\ead{matteo.cinelli@uniroma1.it}
\author[inst1]{Roy Cerqueti}
\ead{roy.cerqueti@uniroma1.it}

\affiliation[inst1]{organization={Department of Social Sciences and Economics, Sapienza University of Rome}, 
addressline={P.le Aldo Moro, 5}, 
city={Rome}, 
postcode={00185}, 
state={Lazio}, 
country={Italy}}

\affiliation[inst2]{organization={Department of Computer Science, Sapienza University of Rome}, 
addressline={Viale Regina Elena, 295}, 
city={Rome}, 
postcode={00161}, 
state={Lazio}, 
country={Italy}}

% Corresponding author
%\cortext[cor1]{Corresponding author. Mobile: +xxxxx}

\begin{abstract}

Detecting multimodality in empirical distributions is a fundamental problem in statistics and data analysis, with applications ranging from clustering to the study of complex systems. In practice, however, assessing departures from unimodality in a consistent and comparable way remains challenging. Widely used methods such as Hartigan and Hartigan’s Dip Test illustrate these difficulties, as the interpretation of their statistics depends strongly on sample size, requires calibration to determine significance, and, for large samples, exhibit increasing sensitivity, leading to rejection of unimodality for arbitrarily small deviations from the null. 
We introduce Z-Dip, a standardized measure of multimodality that addresses these limitations. By treating the Dip statistic as a random variable under the null hypothesis of unimodality and standardizing its observed value, the proposed approach yields scores that are directly comparable across datasets of different sizes. Using simulation-based calibration, we derive a universal decision threshold that closely reproduces classical Dip Test decisions without requiring sample-size-specific adjustments.
Extensive validation on simulated data and on more than 88,000 empirical opinion distributions shows near-perfect agreement with the classical Dip Test while providing a more interpretable and comparable measure of modality. Finally, we propose a downsampling-based correction that mitigates residual sensitivity in extremely large samples. Open-source software and reference tables are provided to facilitate practical adoption.
\end{abstract}


\end{frontmatter}

%\maketitle

\section{Introduction}

Multimodality is a relevant feature of many empirical distributions, reflecting the presence of multiple concentrations of data around specific values, which are often indicative of heterogeneous generating processes or coexisting regimes underlying the observed phenomenon. Correctly identifying departures from unimodality is therefore of broad interest across empirical sciences, both as a descriptive task and as a way to detect nontrivial structure in observed data: as modern applications increasingly rely on large-scale, high-dimensional, and weakly structured data, automated and reliable methods for detecting and quantifying departures from unimodality have become essential to inform decisions \cite{schelling2020dataset, chasani2022uu}.
Examples of multimodal patterns arise across a wide range of domains. In complex networks, the modality of degree distributions informs structural inference \cite{BARUCCA2016244}, distinguishing between homogeneous connectivity and more structured organizations such as core–periphery architectures \cite{paul2004optimization, PhysRevE.71.047101, PhysRevLett.92.118702}. Multimodality has also been observed in other graph-related quantities, including path-length statistics \cite{DONG2025116124} and redundancy measures \cite{jia2013emergence}. Similar considerations apply to social systems, where multimodal distributions of beliefs or stances emerge from interaction and aggregation processes. In this context, bimodality is often used as a proxy for political polarization \cite{Bramson_Grim_Singer_Berger_Sack_Fisher_Flocken_Holman_2017}, and has been linked to fragmented discourse and limited cross-group interaction \cite{10.1093/pnasnexus/pgaf262, Barberá_2015, doi:10.1073/pnas.2023301118}.

Following this perspective, multimodality is not merely a descriptive property of empirical data, but can also be an indicator of structural properties of the system or a description of complex underlying processes. Correctly identifying it, particularly across large collections of samples where visual inspection might be impractical, is therefore an important, and often far from trivial task, which remains the subject of active research~\cite{chasani2025statistical}. 

A prominent example is Hartigan and Hartigan’s Dip Test \cite{10.1214/aos/1176346577}, a widely used nonparametric method for assessing unimodality. Its statistic, known as the Dip, measures the maximal deviation between the empirical cumulative distribution function and the closest unimodal cumulative distribution, obtained via a linear envelope construction. By relying solely on order statistics and shape constraints rather than parametric assumptions, the Dip Test provides a simple and robust way to detect deviations from unimodality. Several alternative approaches can be considered, including Silverman’s kernel density-based test \cite{10.1111/j.2517-6161.1981.tb01155.x}, excess mass tests \cite{10.1111/1467-9868.00141, repec:spr:testjl:v:28:y:2019:i:3:d:10.1007_s11749-018-0611-5}, and mode-tree methods \cite{Minnotte01031993}. While these methods offer complementary perspectives, they typically require tuning parameters, are computationally more demanding, or lack the conceptual immediacy of rank-based approaches. As a result, Dip Test-based methods have been adopted and continue to be widely used across a range of domains, from empirical studies of political behavior~\cite{flamino2023political, loru2025ideology, di2025quantifying}, to feature engineering~\cite{ding2018precision} and deep learning-based clustering \cite{kalogeratos2012dip, leiber2021dip, li2022searching, 10.1145/3534678.3539407, 10.1007/s00521-024-10497-4}.

These methods, however, present a set of limitations that reduce their interpretability, especially in large-scale applications. First, critical values of the Dip statistic are strongly depend on sample size, thus requiring lookup tables or Monte Carlo simulation for assessment of significance. Second, the Dip statistic is not standardized, preventing meaningful comparisons across datasets of different sizes. Third, for large samples, the test rejects unimodality even for distributions that are in fact unimodal, since the null distribution of the Dip statistic narrows as the sample size increases. These issues complicate the use of Dip Test-based methods when distributional comparisons are performed across many subpopulations or time points, or when the goal is not only to test a deviation from unimodality, but to quantify its extent.

In this paper we introduce Z-Dip, a standardized measure of multimodality designed to address these challenges. Rather than treating modality detection purely as a hypothesis-testing problem, we rework the Dip statistic as a standardized score measuring the strength of deviation from unimodality. Specifically, we treat the Dip statistic as a random variable under the null hypothesis of unimodality and standardize its observed value using the mean and variance of its null distribution. This transformation converts the Dip from a sample-size dependent test statistic into a scale-independent measure of multimodality that is directly comparable across datasets and sample sizes. We show that the resulting score has a stable null distribution across a wide range of sample sizes, enabling the use of a single universal threshold for multimodality detection while preserving the original test’s computational complexity and parameter-free formulation. In this way, Z-Dip extends the Dip test from a binary hypothesis test to a continuous, standardized measure of multimodality, while retaining the original decision rule as a simple threshold on the standardized score. 

We validate the Z-Dip on both simulated and empirical data. Simulated tests demonstrate that the Z-Dip reproduces Dip Test decisions across a number of sample sizes while providing a more interpretable scale. An empirical analysis of more than 88${,}$000 opinion distributions shows practically perfect agreement between the Z-Dip threshold and the classical Dip Test. Finally, we present an effective downsampling-based correction that mitigates residual sample-size effects in very large datasets. An open-source R implementation of the Z-Dip is made available at \url{https://github.com/EdoardoDima/zdip}.

The rest of the paper is organized as follows: Section~\ref{sec:hartigan} introduces the theoretical background of Hartigans’ Dip Test and outlines its limitations. We then formally introduce the Z-Dip in Section~\ref{sec:zdip}, and show that the standardized formulation admits a universal significance threshold that closely reproduces the decisions of the classical Dip Test without requiring sample-size-specific calibration. We provide results of its calibration and validation in Section~\ref{sec:validation}, using both simulated and empirical data, and, finally, we discuss our proposed correction for the distortion of these measures at large sample sizes in Section~\ref{sec:downsampling}. Considerations regarding our findings and results can be found in Section~\ref{sec:conclusions}.

\section{Background}
\label{sec:hartigan}
\subsection{Hartigans’ Dip Test}

The Dip Test, introduced by Hartigan and Hartigan in 1985 \cite{10.1214/aos/1176346577}, is a classical nonparametric procedure for testing whether a distribution is unimodal. Given a sample $X_1,\ldots,X_N$ drawn from an unknown distribution with Cumulative Distribution Function (CDF) $F$, the test evaluates the null hypothesis that $F$ is unimodal against the alternative that it is not.

A distribution is unimodal if its density has a single local maximum. Equivalently, its CDF is convex up to the mode and concave thereafter. Let $\mathcal{U}$ denote the class of unimodal distribution functions on $\mathbb{R}$. The Dip Test measures how far the empirical cumulative distribution function $F_N(x)$ deviates from this class.

Formally, the Dip statistic is defined as
\[
\mathrm{Dip}(F_N) = \inf_{G \in \mathcal{U}} \sup_{x} |F_N(x) - G(x)|,
\]
that is, the smallest Kolmogorov–Smirnov distance between the empirical distribution function and the class of unimodal CDFs.

Hartigan and Hartigan showed that this quantity can be computed efficiently by exploiting the shape constraints implied by unimodality. In particular, the algorithm constructs two envelopes around the Empirical Cumulative Distribution Function (ECDF): the greatest convex minorant (GCM), defined as the largest convex function lying below $F_N$, and the least concave majorant (LCM), defined as the smallest concave function lying above $F_N$. These envelopes define candidate unimodal CDFs by joining segments where the slopes satisfy the compatibility condition
\[
\nabla \mathrm{GCM} \le \nabla \mathrm{LCM}.
\]
The Dip statistic corresponds to the maximum vertical distance between the ECDF and the closest admissible unimodal CDF obtained through this construction. In more intuitive terms, the Dip statistic quantifies the smallest perturbation required to transform the empirical distribution into one that satisfies the unimodality constraint, where larger values therefore indicate stronger departures from unimodality. To assess statistical significance, the observed Dip value is compared to its sampling distribution under the null hypothesis of unimodality. Hartigan and Hartigan proposed using the uniform distribution on $[0,1]$ as a reference null, as it represents a least-favorable unimodal distribution. In practice, significance can be evaluated either through Monte Carlo simulation, by generating samples of size $N$ from a uniform distribution and computing their Dip statistics, or by consulting tabulated critical values provided in \cite{10.1214/aos/1176346577}. If the observed Dip exceeds the corresponding critical value, unimodality is rejected.

A key strength of the Dip Test lies in its simplicity: it is fully nonparametric and requires no user-specified smoothing parameters or model assumptions. Unlike kernel-based or mixture-based methods, it can be applied directly to empirical data without tuning or model selection, which has contributed to its wide adoption across many fields.

\subsection{Limitations of the classical Dip Test}

Despite its merits, the classical Dip Test presents several practical and conceptual limitations that hinder its interpretability and use in large-scale applications.
\newline
\begin{enumerate}[label=(\roman*)]
    \item \textbf{Dependence on sample size.} 
    The null distribution of the Dip statistic varies strongly with sample size $N$. As $N$ increases, the Dip values expected under unimodality decrease systematically, since larger samples provide finer-grained approximations to the true underlying distribution, which has a flat density, and thus a linear CDF. Consequently, each sample size requires its own set of critical values or a separate Monte Carlo calibration. This dependence complicates both interpretation and automation, as results from samples of different sizes cannot be directly compared.
       
    \item \textbf{Lack of standardization.}
    The Dip statistic itself is not standardized. Its raw magnitude depends on the scale of sampling variability and on $N$, making it unsuitable for quantitative comparison across datasets. %\mc{mi trovi d'accordo}
    For instance, a Dip of $0.1$ may signal multimodality in a small sample, but be negligible in a large one. Without a standardized scale, the test can only yield binary decisions (reject or not reject unimodality), rather than interpretable effect sizes that quantify ``how multimodal'' a distribution is. However, the Dip statistic already encapsulates this information: it measures the strength of deviation from unimodality, but this signal is obscured by sample-size dependence.

    \item \textbf{Inconsistencies at large sample sizes.}
    As already mentioned, for large $N$, the Dip Test becomes increasingly sensitive to even minor deviations from the ideal unimodal shape. Because the null distribution of the Dip shrinks toward zero as $N$ grows, the smallest irregularities, such as sampling noise or slight skewness, are sufficient to reject unimodality. In practice, this produces false positives in large samples: distributions that are essentially unimodal, but statistically classified as multimodal. This issue means that the Dip Test’s binary decision rule becomes less informative with increasing $N$, effectively turning every large sample into a ``significant" case. As a result, the test loses its ability to distinguish meaningful multimodality (e.g., distinct peaks) from trivial numerical deviations.
\end{enumerate}

\vspace{0.75em}
Together, these limitations point to the need for a standardized, scale-\mc{independent} reformulation of the Dip Test that preserves its parameter-free simplicity while allowing meaningful comparison across sample sizes, maintaining interpretability at large $N$, and eliminating the need for case-specific calibration. The next section introduces the Z-Dip, which addresses these issues by standardizing the Dip relative to its null distribution.

\section{Z-Dip}
\label{sec:zdip}
\subsection{Definition of the Z-Dip Statistic}

We have already introduced how the behavior of the classical Dip statistic is strongly influenced by sample size: as $N$ increases, the empirical cumulative distribution function converges uniformly to the underlying distribution, and the null distribution of the Dip contracts toward zero. Essentially, the Dip under unimodality scales with the resolution of the ECDF, which increases with $N$, and, in consequence, the mean and variance of the null Dip decrease as $N$ grows.

To address this, we propose a standardized version of Hartigan's Dip statistic, which we call the Z-Dip. For an observed dataset of size $N$, let $\text{Dip}(F_N)$ denote its Dip statistic's value. We define

\begin{equation}
Z\text{-}Dip = Z_N = \frac{\text{Dip}(F_N) - \mu_N}{\sigma_N},
\label{eq:zdip}
\end{equation}

where $\mu_N$ and $\sigma_N$ denote the mean and standard deviation of Dip values under the null distribution of $N$-sized samples drawn from the uniform distribution on $[0, 1]$, consistently with the original formulation of the Dip test. 
Standardization, therefore, removes the dominant sample-size effect.
If $\mu_N$ and $\sigma_N$ equal the true null moments, then under unimodality the standardized statistic satisfies
\[
\mathbb{E}_0[Z_N] = 0, 
\qquad 
\mathrm{Var}_0(Z_N) = 1,
\]
so that the null distribution is centered and variance-stabilized across $N$.

Although a full asymptotic characterization of the Dip statistic is beyond the scope of this work, empirical evidence suggests that the standardized statistic exhibits approximate distributional stability across a wide range of sample sizes under unimodality, with minor finite-sample deviations from normality, also explaining why a universal decision threshold can be derived and employed across a wide range of sample sizes. In summary, the score proposed in Formula \eqref{eq:zdip} quantifies how strongly the observed distribution deviates from unimodality, in terms of standard deviation relative to the null expectation. 
Because it rescales the Dip relative to its null, the Z-Dip provides a scale-independent, comparable measure across different sample sizes. Furthermore, while the Z-Dip is inherently unbounded, in contrast to the classical Dip statistic, monotone transformations may be applied if a bounded scale is preferred (Fig. S1 of the Supplementary Information (SI)). 
\subsection{Asymptotic behavior}

The behavior of the Z-Dip can be understood by examining how the magnitude of sampling fluctuations in the empirical distribution scales with the sample size.

Let $X_1,\dots,X_N$ be independent samples drawn from a distribution with cumulative distribution function $F$. The empirical distribution

\[
F_N(x) = \frac{1}{N}\sum_{i = 1}^N \mathbf{1}\{X_i \leq x\}, \qquad x \in \mathbb R
\]

converges uniformly to the true distribution. In particular, the Dvoretzky–Kiefer–Wolfowitz inequality states that for any $\varepsilon > 0$

\begin{equation}
    \label{Cheb}
\Pr\!\left(\sup_{x \in \mathbb R} |F_N(x)-F(x)|>\varepsilon\right)
\le
2e^{-2N\varepsilon^2}.    
\end{equation}

It can also be shown that, by fixing a target error probability $\alpha$ such that $\alpha=\Pr\!\left(\sup_{x \in \mathbb R} |F_N(x)-F(x)|>\varepsilon\right)=2e^{-2N\varepsilon^2}$, solving for the deviation between the ECDF and the CDF returns

\[
\varepsilon = \sqrt{\frac{\log(\frac{2}{\alpha})}{2N}} = 
\sqrt{\frac{\log(\frac{2}{\alpha})}{2}}\cdot N^{-1/2}
\]

implying that the maximum vertical distance between the empirical and true CDF shrinks at rate $O_p(N^{-1/2})$.

\paragraph{Scaling under unimodality}
Given these considerations, first consider the case where the underlying distribution $F$ is unimodal. Since the Dip statistic measures the minimal shift
required to transform the empirical distribution into one that satisfies the unimodality constraint:

\[
\text{Dip}(F_N) = \inf_{G\in U} \sup_{x \in \mathbb R} |F_N(x)-G(x)|,
\]

where $U$ denotes the set of unimodal cumulative distribution functions identified by the algorithm, it follows that, when the underlying distribution $F$ is unimodal, the true distribution itself belongs to the admissible class $U$. Consequently,

\[
\text{Dip}(F_N) \le \sup_{x \in \mathbb R} |F_N(x)-F(x)|.
\]

Since the empirical deviation from the CDF scales as $O_p(N^{-1/2})$, the Dip statistic must also shrink at least at this rate. This was already shown by Hartigan and Hartigan in their original paper, where it was noted how $\mathrm{Dip}(F_N) \rightarrow 0$. At the same time, given the Z-Dip formulation: 

\[
Z_N = \frac{\text{Dip}(F_N) - \mu_N}{\sigma_N},
\]

the standard deviation of the Dip statistic under the null distribution is driven by the same process, resulting in both the numerator and denominator to scale at the same rate in $N$, leading to a Z-Dip distribution that remains stable around $0$ across sample sizes.

\paragraph{Scaling under multimodality}
If instead the underlying distribution $F$ is not unimodal, it will not be part of the set $U$ of candidate unimodal distributions. 

Let us consider $\tilde G \in U$ and $\bar x \in \mathbb R$ such that

\[
\text{Dip}(F_N) =  |F_N(\bar x)-\tilde{G}(\bar x)|.
\]
There exists a positive constant $c=c(\tilde G, \bar x)$ such that
\[
 c =  |F(\bar x)-\tilde{G}(\bar x)|,
\]

and one simply has:

\[
{\rm{Dip}} (F_N)= |F_N(\bar x)-\tilde{G}(\bar x)| = |F_N(\bar x)-F(\bar x) + F(\bar x)-\tilde{G}(\bar x)|.
\]

Since the empirical CDF converges to $F$, the right-hand side of the inequality asymptotically vanishes, implying:

\[
\mathrm{Dip}(F_N) \rightarrow c,
\]

meaning that, for distributions that deviate from unimodality, the Dip statistic converges to a positive constant $c$ reflecting the distance between the true distribution and the class of unimodal distributions. Substituting this property into our statistic and being $\mu_N \to 0 $ as $N \to \infty$, we have 

\[
Z_N = \frac{\mathrm{Dip}(F_N) - \mu_N}{\sigma_N}
\approx
\frac{c}{\sigma_N}
\quad \text{as } N \to \infty.
\]

We observe that the numerator converges asymptotically to a positive constant, while the standard deviation $\sigma_N$ continues to shrink proportionally to $N^{-1/2}$ by the Chebychev inequality and according with Formula \eqref{Cheb}. Consequently, 

\[
Z_N \sim \frac{c}{N^{-1/2}} \propto N^{1/2},
\]

explaining its empirical power-law behavior for multimodal cases.

\paragraph{Empirical validation of asymptotic assumptions}

\begin{figure}[H]
 \centering
    \includegraphics[width=.9\linewidth]{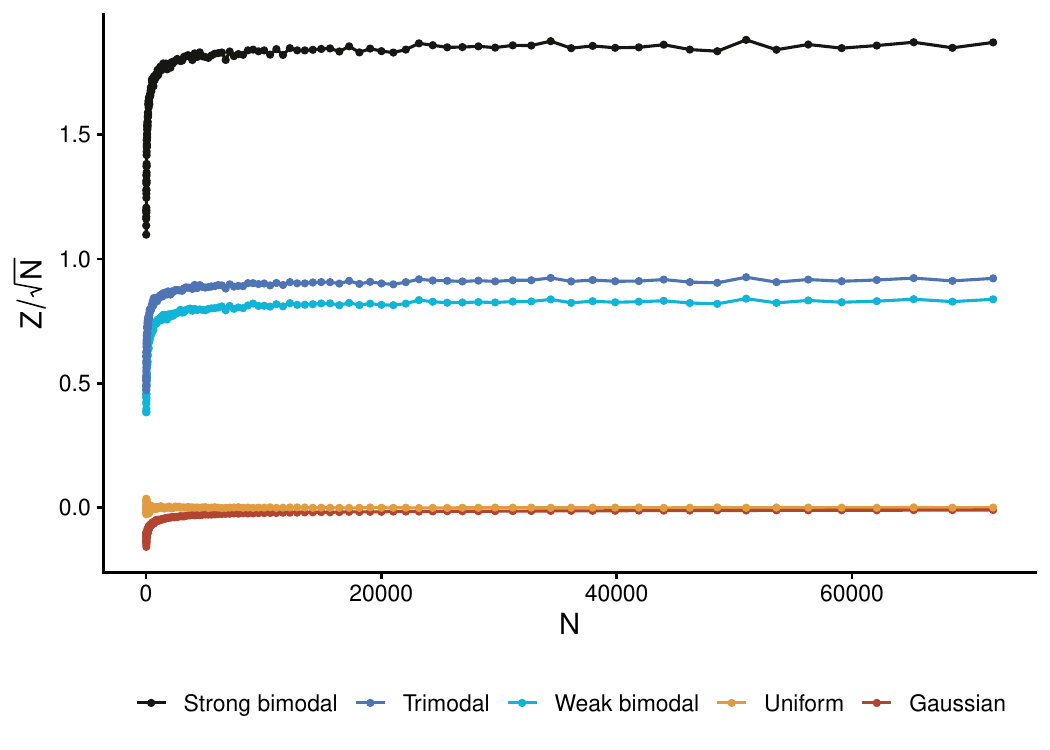}
    \caption{Scaling of Z-Dip across modality regimes. Under unimodality, the rescaled statistic $Z_N /N^{1/2}$ converges to zero. Under multimodality, it stabilizes to a positive constant, consistent with the theoretical scaling $Z_N \propto N^{1/2}$. Plateau differences indicate varying strengths of multimodal structure. 
    }
    \label{fig:scaling_N}
\end{figure}

These asymptotic considerations are also supported by empirical simulation results, as shown in Figure~\ref{fig:scaling_N}. Under the proposed asymptotic regime described above, if the underlying distribution deviates from unimodality, then the Dip statistic converges to a positive constant $c$, while the null standard deviation scales as $N^{-1/2}$; thus,

\[
\frac{Z_N}{N^{1/2}} \sim c,
\]

This behavior is clearly observed in Figure~\ref{fig:scaling_N}, where, for unimodal distributions (uniform and gaussian), the rescaled statistic $Z_N / N^{1/2}$ converges to zero, indicating that no persistent structural deviation is present and the observed values are driven entirely by sampling noise. In contrast, for multimodal distributions $Z_N /N^{1/2}$ rapidly stabilizes to a positive constant, where the different plateau levels correspond to different values of $c$, capturing the strength of the underlying multimodal structure. Moreover, the fact that $Z_N / N^{1/2}$ converges to a constant confirms that $Z_N$ grows proportionally to $N^{1/2}$.

\newpage

\subsection{Simulation of Null Distributions}

To estimate $\mu_N$ and $\sigma_N$, we simulate large ensembles ($20,000$ iterations each) of null samples from the uniform distribution for a grid of 183 roughly log-spaced sample sizes $N$, ranging from small ($N=4$) to very large ($N=72{,}000$). The uniform distribution is used in line with Hartigan and Hartigan's original formulation, where it was used to calibrate empirical percentage points of the Dip statistic. For each simulated dataset, we compute the Dip statistic, yielding an empirical null distribution, and the mean and standard deviation of this distribution provide estimates of $\mu_N$ and $\sigma_N$. 

For sample sizes not explicitly simulated, $\mu_N$ and $\sigma_N$ are obtained by linear interpolation between adjacent grid points, and, to enable efficient application, these values are stored in precomputed lookup tables, ensuring that Z-Dip scores can be rapidly obtained for any sample size without repeated simulation. The binary files containing the complete lookup tables are provided with the software accompanying this paper (see Section \ref{sec:data_code}).

\section{Results}
\label{sec:validation}
\subsection{Comparison with the Classical Dip Statistic}

\begin{figure}[h]
 \centering
    \includegraphics[width=\linewidth]{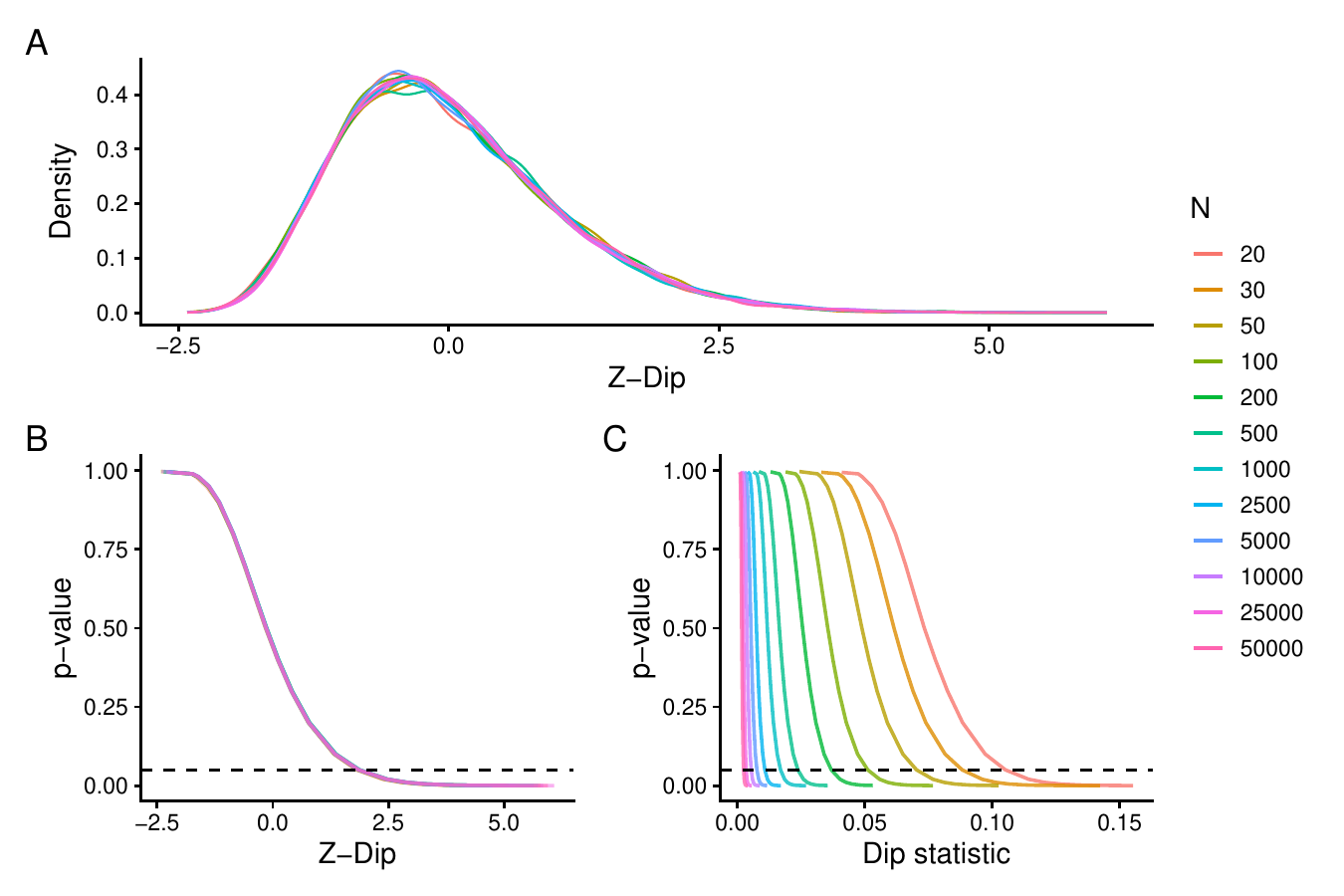}
    \caption{Comparison of the classical Dip statistic and the standardized Z-Dip across sample sizes. Z-Dip distributions were obtained over 20${,}$000 simulated uniform samples.
    \textbf{(A)} Density of Z-Dip scores for different sample sizes ($N$), showing strong overlap of null distributions, proving that standardization removes the dependence on $N$. 
    \textbf{(B, C)} Lineplot of Z-Dip and Dip statistic values versus corresponding Dip Test $p$-values for simulated unimodal distributions, colored by sample size $N$.
    }
    \label{fig:fig1}
\end{figure}

We first compare the behavior of the raw Dip and the standardized Z-Dip across sample sizes. The classical Dip values show strong sample-size dependence: as $N$ increases, the expected Dip under the null shrinks toward zero, while its variance decreases, making large-sample values numerically small and difficult to interpret. By contrast, the Z-Dip eliminates this dependence. Across all simulated sample sizes, the null distribution of the Z-Dip is centered around zero with approximately constant spread (Figure~\ref{fig:fig1}, panel A). This confirms that standardization removes the sample-size dependence inherent in the raw Dip statistic.

A direct comparison between Dip $p$-values and corresponding Z-Dip scores (Figure~\ref{fig:fig1}, panel B) further demonstrates the strong monotonic relationship between the two measures. This alignment suggests that a single Z-Dip threshold can reproduce conventional significance levels across all $N$, in contrast with the regular Dip statistic (Figure~\ref{fig:fig1}, panel C).

\subsection{Threshold Calibration}

Under the null hypothesis of unimodality, the Z-Dip is approximately normally distributed with mean zero and unit variance. If this normal approximation were exact, a two-tailed 5\% significance level would correspond to the canonical threshold $z = 1.96$. However, empirical simulations reveal that the null distribution of the Z-Dip values is slightly left-skewed (Fig. \ref{fig:fig1}, panel A). Even after standardization, which enforces centering and variance normalization, but does not impose Gaussianity, minor deviations from normality persist, likely due to the inherent skewed nature of the classical Dip statistic. 
As a result, adopting the theoretical $z = 1.96$ cutoff could yield false-negative rates marginally above the 5\% level.

To ensure robustness across sample sizes, we estimate the Z-Dip threshold corresponding to $p = 0.05$ empirically. For each $N \ge 20$, we compute Dip $p$-values and corresponding Z-Dip scores for a large ensemble of null samples, then identify the Z-Dip value at which $p = 0.05$ via linear interpolation. Bootstrap resampling provides confidence intervals for each estimated threshold. These values are highly consistent across $N$, with upper bounds ranging narrowly around $1.85$, indicating that the residual dependence of the standardized statistic on sample size is minor. We therefore adopt a conservative universal cutoff $z = 1.85$, corresponding closely to the theoretical $1.96$ threshold while ensuring that the nominal Type I error rate is not exceeded for any sample size. The choice to calibrate the threshold only on samples having $N \ge 20$ is due to the greater variability of the scores at lower sample sizes, where the coarser approximation of the empirical distribution and the reduced resolution of the Dip statistic lead to more unstable Z-Dip values and less reliable threshold calibration.

\subsection{Computational Considerations}

The computation of the Dip statistic, as originally defined by Hartigan and Hartigan, involves finding the maximum deviation between the empirical distribution function and the closest unimodal distribution. The algorithm proceeds by sorting the observed data, constructing candidate unimodal envelopes, and computing the maximal deviation. The sorting step dominates the computational cost, leading to an overall time complexity of $O(n \log n)$, where 
$n$ is the number of items to be sorted. After sorting, the iterative identification of modal intervals and the computation of cumulative deviations are linear in $n$, so the asymptotic complexity is determined by the sorting operation.

The Z-Dip retains this computational structure. Once the Dip statistic $Dip_{obs}$ is computed, the standardized Z-dip score, defined in Formula \eqref{eq:zdip}, is obtained by a constant number of arithmetic operations, which have $O(1)$ complexity. Therefore, the total runtime of the Z-Dip remains $O(n \log n)$. Crucially, the Z-Dip does not require repeated Monte Carlo sampling or permutation testing to obtain significance levels. Once lookup tables for $\mu_N$ and $\sigma_N$ are precomputed, evaluating the Z-Dip for new samples becomes effectively instantaneous, making it suitable for large-scale applications, such as distributional analysis across thousands of distributions \citep{di2025quantifying}.

\subsection{Synthetic Testing}

We validated the Z-Dip statistic on controlled synthetic distributions with known modality structure. Synthetic samples were generated using a Gaussian mixture simulator of the form

\[
y_i \sim \sum_{k=1}^{K} \pi_k \, \mathcal{N}(\mu_k, \sigma_k^2),
\]
where $K$ is the number of modes, and
\[
\boldsymbol{\mu} = (\mu_1, \dots, \mu_K), \quad 
\boldsymbol{\sigma} = (\sigma_1, \dots, \sigma_K), \quad 
\boldsymbol{\pi} = (\pi_1, \dots, \pi_K)
\]
denote respectively the vectors of means, standard deviations, and mixing proportions of the $K$ Gaussian components, with $\pi_k \ge 0$ \mc{$\forall \, k$} and $\sum_{k=1}^{K} \pi_k = 1$. The generator function allows full control over the number, separation, and spread of modes, enabling testing from clearly unimodal to strongly bimodal or multimodal cases. For each configuration, we simulated 280 sample sizes $N$, logarithmically spaced between 20 and 5${,}$000, and for each $N$ performed 1${,}$000 independent replications. For each replication, we computed both the classical Dip Test $p$-values and the corresponding Z-Dip scores.

Across all simulations, the Z-Dip preserved the inferential behavior of the Dip Test, with \mc{a nearly perfect} correspondence between significance decisions: across 1${,}$120${,}$000 samples, only in 110 cases the two measures were not aligned in their classification, a discrepancy easily attributable to numerical approximations and/or boundary cases. The specific data-generating processes were defined as follows:

\vspace{0.5em}
\begin{itemize}
    \item \textbf{Unimodal:} ${\mu} = (0)$, ${\sigma} = (0.1)$
    \item \textbf{Weak bimodal:} ${\mu} = (-0.6, 0.6)$, ${\sigma} = (0.15, 0.15)$, ${\pi} = (0.7, 0.3)$
    \item \textbf{Strong bimodal:} ${\mu} = (-0.6, 0.6)$, ${\sigma} = (0.1, 0.1)$, ${\pi} = (0.5, 0.5)$
    \item \textbf{Trimodal:} ${\mu} = (-0.75, 0, 0.75)$, ${\sigma} = (0.1, 0.1, 0.1)$, ${\pi} = (1/3, 1/3, 1/3)$
\end{itemize}
\vspace{0.5em}

Note how the unimodal case simply represents samples drawn from a normal distribution, and that in the in weak bimodal case, we provide $\sigma_1 = \sigma_2 = 0.15$, instead of $\sigma = 0.1$ used in the other cases, to make the modes slightly less peaked and prominent.

Table~\ref{tab:synthetic_validation} summarizes the results across unimodal, weak bimodal, strong bimodal, and trimodal conditions. Multimodal mixtures yielded large positive Z-Dip values and were consistently detected as such, while unimodal distributions remained centered near zero or negative values. It can also be noted that, for genuinely multimodal distributions, the Z-Dip tends to grow systematically with sample size $N$. This behavior arises because well-separated modes produce a persistent ``dip" in the empirical cumulative distribution function: as $N$ increases, the separation between peaks is resolved more clearly, leading to larger Dip statistics and correspondingly higher Z-Dip scores. This behavior is further illustrated in Fig.~S2 in the SI. In contrast, unimodal distributions exhibit flat or slowly varying Z-Dip values with $N$, since noise dominates and no persistent secondary mode exists. These synthetic results provide a controlled validation, showing that the Z-Dip behaves as theoretically expected under both null (unimodal) and alternative (multimodal) conditions.

\begin{table}[h]
\centering
\caption{Synthetic validation of the Z-Dip statistic under varying modality and sample size (1${,}$000 replications per each combination of $N$ and configuration).}
\label{tab:synthetic_validation}
\begin{adjustbox}{width=\columnwidth}
\begin{tabular}{ccccccc}
%\small
\toprule
Configuration & $N$ (range) & Mean Z-Dip & $p<0.05$ (Dip Test) & $Z>1.85$ & \# of simulations & Agreement \\
\midrule
\addlinespace
Unimodal & 20-50 & -0.750 & 0.003 & 0.003 & 31${,}$000 & 1.000\\
 & 51-100 & -0.952 & 0.000 & 0.000 & 37${,}$000 & 1.000\\
 & 101-500 & -1.253 & 0.000 & 0.000 & 87${,}$000 & 1.000\\
 & 501-1000 & -1.547 & 0.000 & 0.000 & 37${,}$000 & 1.000\\
 & $>$1000 & -1.835 & 0.000 & 0.000 & 88${,}$000 & 1.000\\
\addlinespace
Weak bimodal & 20-50 & 5.515 & 0.971 & 0.971 & 31${,}$000 & 0.999\\
 & 51-100 & 9.180 & 1.000 & 1.000 & 37${,}$000 & 1.000\\
 & 101-500 & 19.000 & 1.000 & 1.000 & 87${,}$000 & 1.000\\
 & 501-1000 & 35.434 & 1.000 & 1.000 & 37${,}$000 & 1.000\\
 & $>$1000 & 67.489 & 1.000 & 1.000 & 88${,}$000 & 1.000\\
\addlinespace
Strong bimodal & 20-50 & 7.517 & 0.994 & 0.994 & 31${,}$000 & 1.000\\
 & 51-100 & 12.320 & 1.000 & 1.000 & 37${,}$000 & 1.000\\
 & 101-500 & 25.001 & 1.000 & 1.000 & 87${,}$000 & 1.000\\
 & 501-1000 & 46.149 & 1.000 & 1.000 & 37${,}$000 & 1.000\\
 & $>$1000 & 87.221 & 1.000 & 1.000 & 88${,}$000 & 1.000\\
\addlinespace
Trimodal & 20-50 & 3.459 & 0.900 & 0.897 & 31${,}$000 & 0.996\\
 & 51-100 & 5.724 & 0.999 & 0.999 & 37${,}$000 & 1.000\\
 & 101-500 & 11.850 & 1.000 & 1.000 & 87${,}$000 & 1.000\\
 & 501-1000 & 22.186 & 1.000 & 1.000 & 37${,}$000 & 1.000\\
 & $>$1000 & 42.414 & 1.000 & 1.000 & 88${,}$000 & 1.000\\
\bottomrule
\end{tabular}
\end{adjustbox}
\end{table}

\subsection{Empirical Validation}

To evaluate the empirical validity of the proposed Z-Dip statistic, we applied it to a large-scale dataset of $88{,}528$ observed opinion distributions having sample size $N \ge 20$. Each distribution represents the political leaning of YouTube users engaging in political video discussions. User leanings are continuous in the range $[-1, +1]$, where $-1$ indicates strongly left-leaning and $+1$ strongly right-leaning users. These distributions were previously constructed and described in~\cite{di2025quantifying}, and provide a rich and heterogeneous collection of empirical distributions, enabling large-scale testing of modality detection methods. 
For each distribution, we computed both the classical Dip statistic and the standardized Z-Dip, using the threshold $z = 1.85$ for significance derived in Section \ref{sec:zdip}.

\begin{figure}[H]
    \centering
    \includegraphics[width=.8\linewidth]{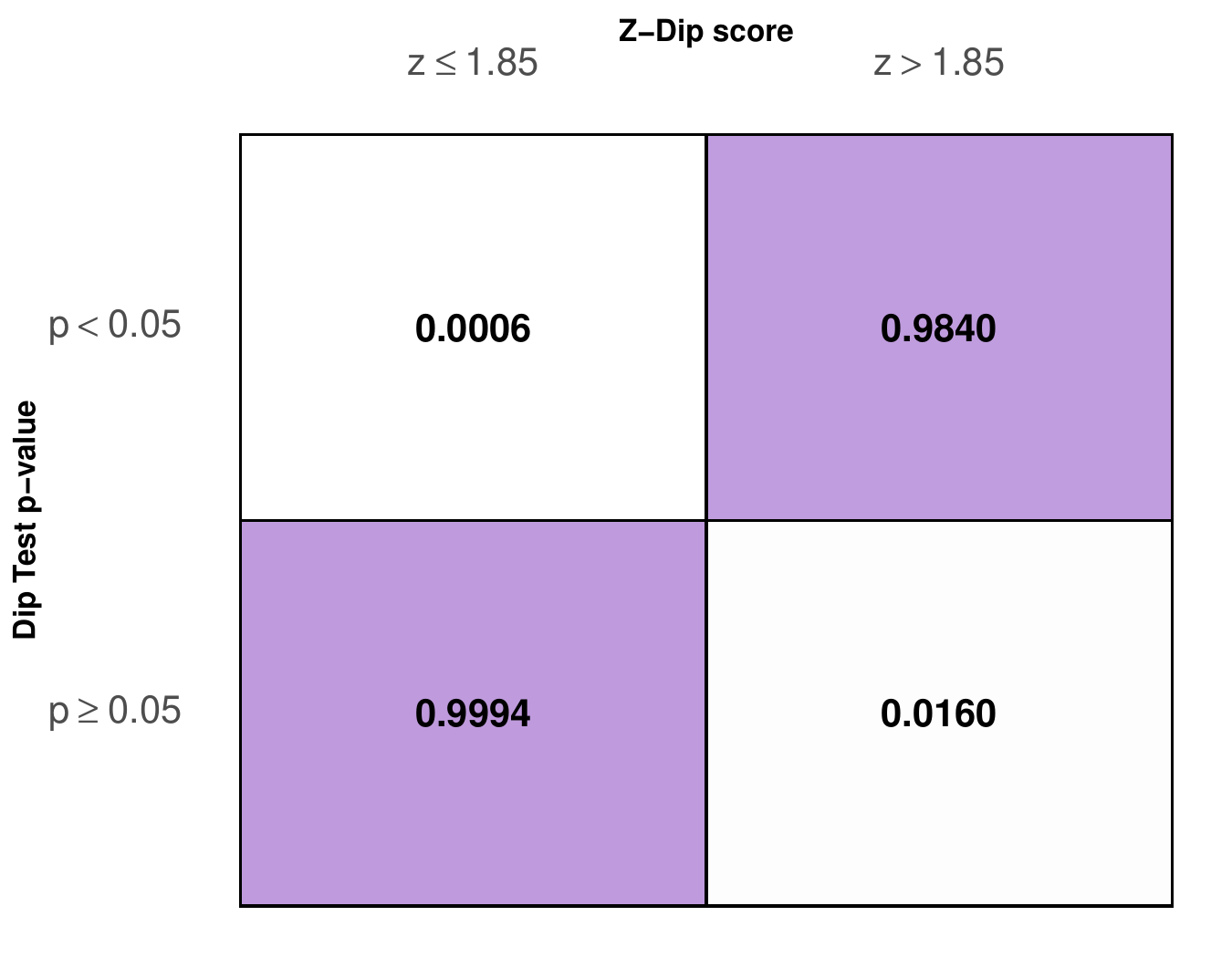}
    \caption{Column-normalized confusion matrix of the agreement between the Z-Dip threshold and the Dip Test's $p$-value over the $88{,}528$ empirical samples considered.}
    \label{fig:fig2}
\end{figure}

Despite the empirical sample being heavily skewed toward unimodal (i.e., non-multimodal) cases, the two methods produced nearly identical classifications, with an overall agreement of $99.899\%$. Moreover, the two measures yield a Cohen's Kappa coefficient of $\kappa = 0.981$, indicating near-perfect agreement that is well above what would be expected by chance. This close correspondence confirms that the Z-Dip effectively reproduces the inferential behavior of the original Dip Test while providing a standardized, comparable measure of deviation from unimodality. Fig.~\ref{fig:fig2} further illustrates these results via a column-normalized confusion matrix between the two measures.

The few discrepancies (89 cases in total) often occur at lower sample sizes. The median number of observations among the disagreeing cases is $N = 89$. This likely stems from the fact that when $N$ is small, the empirical distribution contains too few distinct values to approximate a smooth shape, leading to discrete artifacts and unstable estimates. Under these conditions, small numerical variation, such as rounding differences or the interpolation of $\mu_N$ and $\sigma_N$ in the Z-Dip, can flip the classification outcome. It is also worth noting that a fixed universal threshold ($z = 1.85$) necessarily ignores the slight $N$-dependence of the Dip’s true critical values. Minor disagreements are therefore expected in boundary cases. Nonetheless, across our empirical validation, the agreement between the Z-Dip and the classical test is essentially perfect.

In summary, the Z-Dip preserves the decision structure of Hartigans' Dip Test and generalizes it into a form that is computationally efficient, interpretable, and robust across datasets of varying sizes. Furthermore, the Z-Dip can be treated as a drop-in replacement for the classical Dip Test that eliminates the need for sample-size-specific calibration without sacrificing accuracy or interpretability.

%%%%

%%%%%

\section{Addressing Large-Sample Sensitivity}
\label{sec:downsampling}

A limitation of the classical Dip Test, and by extension the Z-Dip, is its increasing sensitivity at very large sample sizes. As $N$ grows, the null distribution of the Dip statistic contracts toward zero, so even minor irregularities (such as slight skewness or minimal outliers) can produce statistically significant deviations. This leads to a proliferation of practically negligible deviations that are treated as significant, where effectively unimodal distributions are classified as multimodal.

While the Z-Dip standardization improves comparability across sample sizes, the fundamental sensitivity to tiny deviations in large-$N$ samples remains.

One straightforward solution is to resample or downsample the data to a moderate effective sample size before computing the Dip statistic or the corresponding Z-Dip. By reducing $N$ to a size where the null distribution retains meaningful variance, the test becomes less sensitive to trivial deviations while preserving the ability to detect substantive multimodality. 
Concretely, for a large sample of size $N$, one can compute the Z-Dip for $n_{sim}$ times on random subsets of size $N_\text{sub} \ll N$, and then average the resulting Z-Dip scores:

\[
\overline{Z\text{-Dip}} = \frac{1}{n_{\text{sim}}} \sum_{i=1}^{n_{\text{sim}}} Z\text{-Dip}(x_i),
\]

where $x_i$ is a random subset of size $N_\text{sub}$, and $Z\text{-Dip}(x_i)$ is the Z-Dip score computed on the subset $x_i$. This procedure preserves the nonparametric, tuning-free nature of the test, while mitigating the pathological sensitivity of the statistic in very large samples. This downsampling-based approach enables meaningful comparisons across very large datasets and ensures that the test captures substantive deviations from unimodality rather than trivial sampling noise, adding only very little computational overhead.

\begin{figure}[H]
    \centering
    \includegraphics[width=.9\linewidth]{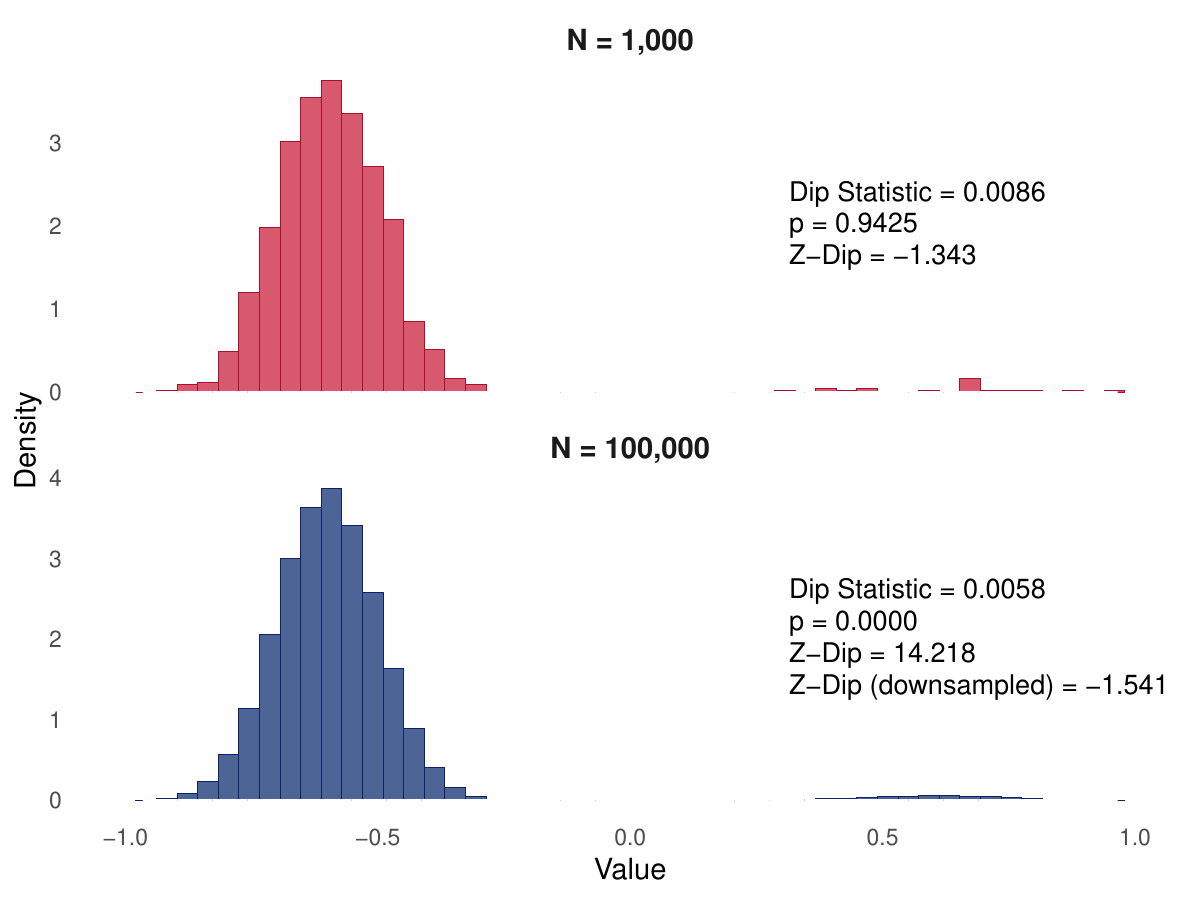}
       \caption{Effect of sample size on Dip Test results and the corresponding downsampled Z-Dip correction. Both samples in the figure contain a detectable but practically negligible secondary mode around $X = 0.5$, which leads the test to classify the large sample as multimodal. The downsampled Z-Dip, however, correctly identifies the distribution as unimodal, illustrating its robustness to scale-induced artifacts.
       }
    \label{fig:fig3}
\end{figure}

Figure~\ref{fig:fig3} illustrates this effect: for two visually similar distributions, one with $N = 1{,}000$ and another with $N = 100{,}000$, with both having a negligible secondary peak further from the mode, the classical Dip Test, along with its $p$-value and the corresponding Z-Dip score, classify the smaller sample as unimodal, and the larger sample as multimodal. In contrast, the downsampled Z-Dip correctly identifies both distributions as unimodal ($z < 1.85)$, demonstrating that it effectively controls for scale-dependent artifacts and preserves meaningful inference about the true modality of the data.

\begin{figure}[H]
    \centering
    \includegraphics[width=\linewidth]{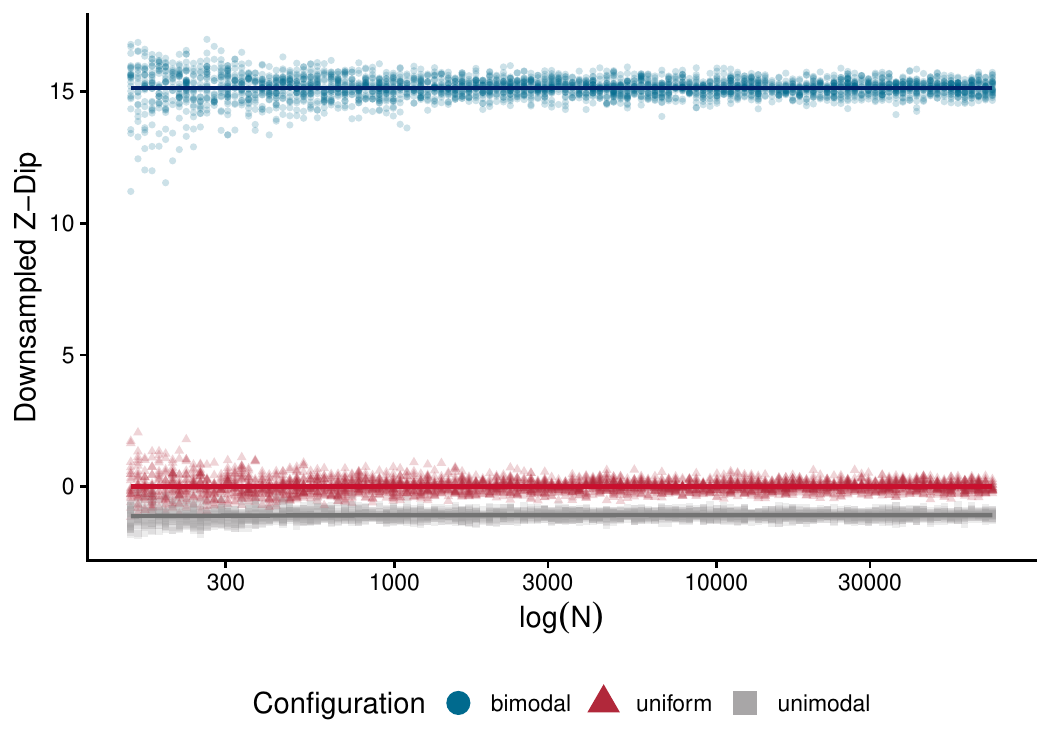}
       \caption{Stable trend of downsampled Z-Dip for sample sizes $N$ ranging from 150 to 72${,}$000, computed for data drawn from three distributions: 
    (i) a bimodal Gaussian mixture with $\mu = (-0.6, 0.6)$, $\sigma = (0.1, 0.1)$, and equal component weights; 
    (ii) a weak bimodal distribution with $\mu = (-0.6, 0.6)$, $\sigma = (0.1, 0.2)$, and a minor mode weight of $0.025$; and 
    (iii) a uniform distribution on $[0, 1]$. 
    Each point represents the mean downsampled Z-Dip over $n_{sim} = 30$.}
    \label{fig:fig4}
\end{figure}

While simple, this procedure remains stable across sample sizes and distributional conditions. To verify this, we conducted a simulation using 132 approximately log-spaced sample sizes, ranging from $N = 150$ to $N = 72{,}000$, to ensure uniform coverage on a multiplicative scale. Samples were drawn from the same underlying distributions (unimodal, bimodal, and uniform).
For each case, we repeatedly computed the downsampled Z-Dip using fixed subsamples of size $N_\text{sub} = 100$. Even for a small number of 30 iterations of subsampling, which adds little computational overhead, the resulting values remained highly consistent across all tested $N$, indicating that the downsampling-based correction neutralizes the scale-dependent inflation observed in the classical Dip and Z-Dip statistics. Figure~\ref{fig:fig4} illustrates this stability, showing that the downsampled Z-Dip values fluctuate minimally even as the total sample size increases by orders of magnitude.

In principle, other strategies could be applied, such as modeling the scaling parameter of the ECDF, or incorporating a sample-size-adjusted penalty term. However, we note that even this simple downsampling approach is sufficient to restore interpretability in high-$N$ scenarios.

\section{Conclusions}
\label{sec:conclusions}
This work introduced the Z-Dip, a standardized reformulation of Hartigans’ Dip Test that enables consistent and interpretable modality assessment across sample sizes. By expressing the Dip in standardized units relative to its null distribution, the Z-Dip transforms a size-dependent hypothesis test into a scale-free descriptive measure. The proposed framework retains the theoretical foundation and computational efficiency of the original Dip Test, while eliminating the need for lookup tables and enabling direct comparison of multimodality strength across datasets.

Validation on synthetic samples and empirical data demonstrate that the Z-Dip preserves the inferential behavior of the Dip Test, achieving virtually perfect correspondence in significance decisions. At the same time, the standardized scale provides additional interpretability, allowing researchers to quantify how strongly a distribution deviates from unimodality. Finally, we proposed a downsampling-based correction to stabilize Z-Dip values in very large samples, enabling robust comparisons across datasets of varying size.

Together, these contributions establish the Z-Dip as a practical, transparent, and computationally efficient tool for detecting and quantifying multimodality. We believe that the availability of reference tables and open-source implementations will facilitate its adoption in applications in a wide range of fields.

\section*{Data and code availability}
\label{sec:data_code}
An open-source R implementation of the Z-Dip is available at \url{https://github.com/EdoardoDima/zdip}. 
The package can be installed locally by following the instructions provided in the repository.
The data used in the empirical validation are available at \url{https://osf.io/vh9px/}. 
Further details on the inference of users' political leaning in the YouTube videos analyzed can be found in \cite{di2025quantifying}.

\section*{CRediT authorship contribution statement}

\textbf{Edoardo Di Martino:} Conceptualization, Methodology, Software, Validation, Formal analysis, Investigation, Data Curation, Visualization, Writing - Original Draft, Writing - Review \& Editing,
\textbf{Matteo Cinelli:} Conceptualization, Supervision, Visualization, Project administration, Writing - Review \& Editing, 
\textbf{Roy Cerqueti:} Conceptualization, Supervision, Project administration, Writing - Review \& Editing,

%\vspace{2em}
\clearpage

%\begin{thebibliography}{5}
%\bibliographystyle{unsrt}
%\singlespacing
%\bibliography{biblio_new}
%\end{thebibliography}
%\end{document}

%\clearpage

%\end{document}

\clearpage

\newcommand{\beginsupplement}{
    \setcounter{section}{0}
    \renewcommand{\thesection}{}
     \setcounter{subsection}{0}
    \renewcommand{\thesubsection}{}
    \setcounter{equation}{0}
    \renewcommand{\theequation}{S\arabic{equation}}
    \setcounter{table}{0}
    \renewcommand{\thetable}{S\arabic{table}}
    \setcounter{figure}{0}
    \renewcommand{\thefigure}{S\arabic{figure}}
    \newcounter{SIfig}
    \renewcommand{\theSIfig}{S\arabic{SIfig}}}
    \setcounter{page}{1}

\beginsupplement

\section*{\centering Supplementary Information for Z-Dip: a validated generalization of the Dip Test for data modality assessment}

%\section*{SI Appendix Section}
\vspace{1.5em}

\subsection*{\textbf{Rescaling the Z-Dip to a fixed scale}}

\begin{figure}[H]
    \centering
    \includegraphics[width=\linewidth]{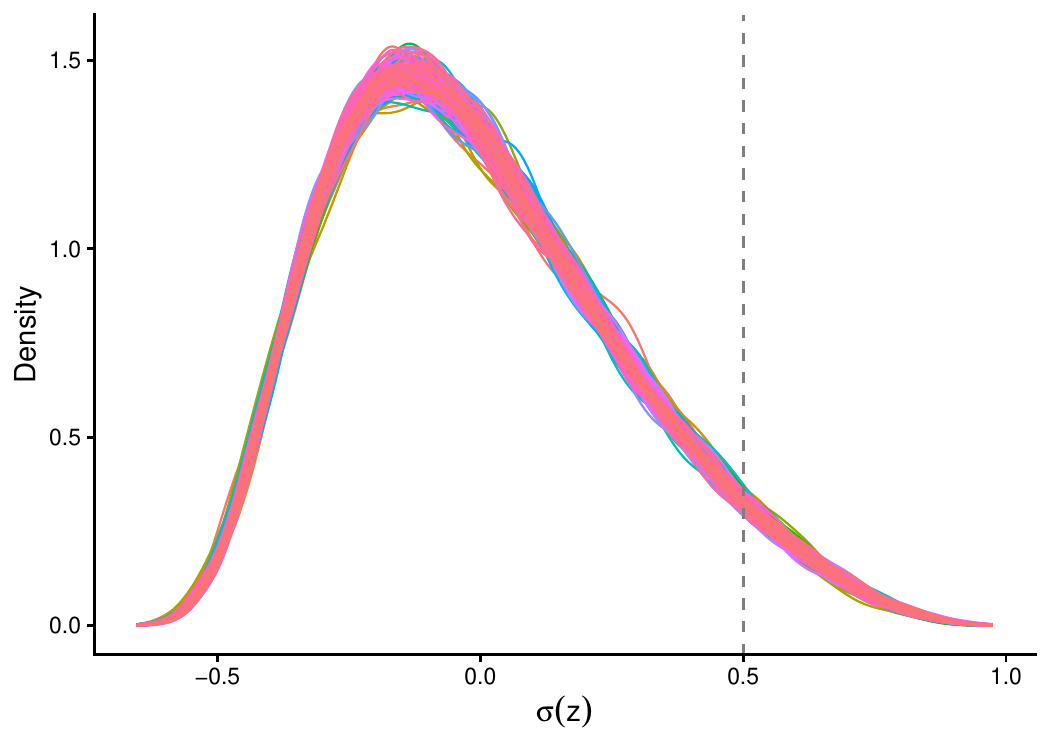}
       \caption{Distributions of Z-Dip scores after applying a sigmoid transformation with $\alpha = 0.595$. The dashed line shows the rescaled significance threshold $\sigma(z) = 0.5$.}
    \label{fig:figs1}
\end{figure}

As mentioned in the main manuscript, the Z-Dip expresses how strongly the observed distribution deviates from unimodality in terms of standard deviation relative to the null expectation. While intuitive, this measure remains unbounded by definition, and we acknowledge how limiting the scores to a certain interval could make the measure more interpretable in certain applications. To do this, a number of squashing functions could be used to map the scores in a certain interval: we show as an example a sigmoid-based function, which remaps scores in the interval $[-1, +1]$:
\[
\sigma(z) = \frac{2}{1 + e^{-\alpha z}} - 1,
\]
for which we propose as a standard $\alpha = 0.595$, the choice of which neatly remaps the threshold $z = 1.85$ to an intuitive $\sigma(z) = 0.5$. Figure \ref{fig:figs1} shows the distribution of the rescaled scores computed on $530{,}000$ uniform samples with sample size ranging from $N=4$ to $N=72{,}000$.

%\clearpage

\subsection*{\textbf{Power-law like scaling of Z-Dip for multimodal distributions}}

\begin{figure}[H]
    \centering
    \includegraphics[width=\linewidth]{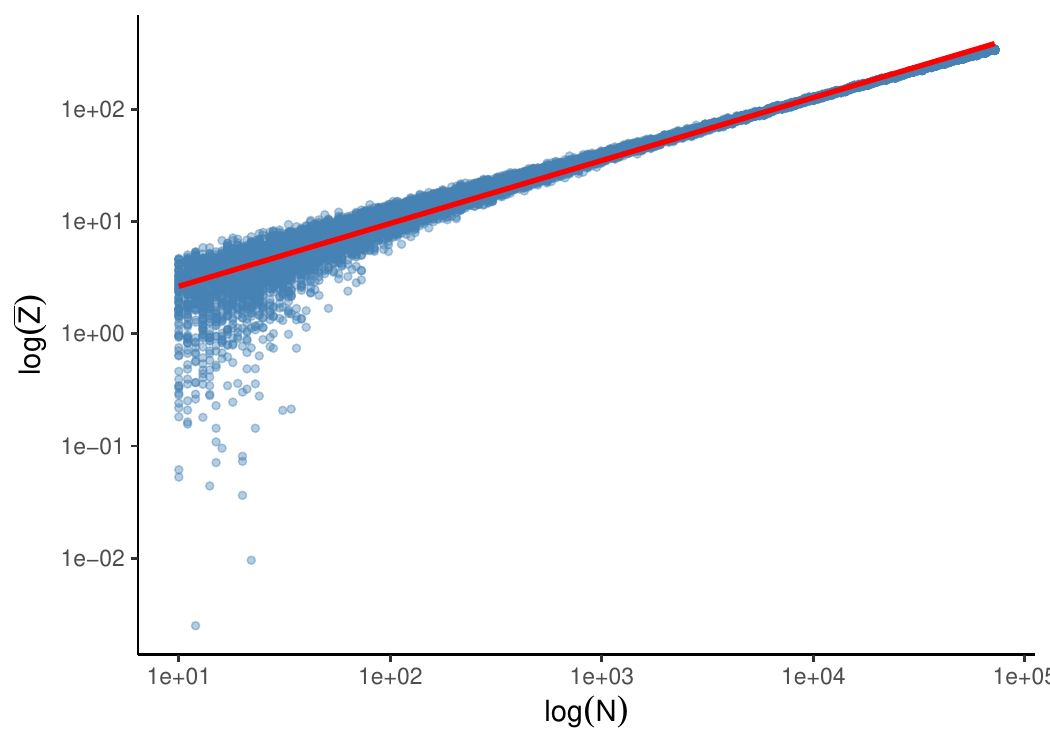}
       \caption{Scaling of Z-Dip as $N$ grows, for explicitly multimodal samples. The average Z-Dip values were computed over 100 samples coming from the same underlying distribution, for each $N$.}
    \label{fig:figs2}
\end{figure}

As introduced in Table~\ref{tab:synthetic_validation} of the main manuscript, for multimodal samples generated from the same underlying distribution, Z-Dip values follow an approximately power-law behavior as a function of $N$ (Figure~\ref{fig:figs2}), such that:

\[
E[\text{Z-Dip}(N)] \propto N^\alpha
\]

\end{document}